\documentclass{jetpl}

\usepackage[cp1251]{inputenc}
\usepackage[english]{babel}

\twocolumn

\def\({\left(}
\def\){\right)}
\def\[{\left[}
\def\]{\right]}

\newcommand{\lr}[1]{ \left( #1 \right) }
\newcommand{\lrs}[1]{ \left[ #1 \right] }

\newcommand{\vev}[1]{ \langle \, #1 \, \rangle }

\newcommand{\expa}[1]{ \exp{\left( #1 \right)} }

\title{The influence of defects on the conductivity of graphene within the effective theory approach}

\rtitle{The influence of defects on the conductivity of graphene\ldots}

\sodtitle{Numerical simulation of graphene in a magnetic field within the effective theory approach}

\author{ S.\,N. Valgushev$^{a,b}$, E.\,V. Luschevskaya$^{a}$, O.\,V. Pavlovsky$^{a,c}$, M.\,I. Polikarpov$^{a, b}$, M.\,V. Ulybyshev$^{a,c}$}

\rauthor{ S.\,N. Valgushev, E.\,V. Luschevskaya, O.\,V. Pavlovsky,
M.\,I. Polikarpov, M.\,V. Ulybyshev}

\sodauthor{Valgushev, Luschevskaya, Pavlovsky, Polikarpov, Ulybyshev}

\address{$^a$Institute of Theoretical and Experimental Physics, 117218 Moscow, Russia}
\address{$^b$Moscow Institute of Physics and Technology, 141700 Dolgoprudniy, Russia}
\address{$^c$Moscow State University, 119899 Moscow, Russia}

\abstract{The results of the simulations by Monte Carlo method of graphene with structural defects are presented.
 The calculations are performed within an effective quantum field theory with
non-compact $3\hm + 1$--dimensional Abelian gauge field and
$2\hm + 1$--dimensional  Kogut-Susskind fermions. It was found that
defects shift the phase transition point semimetal-insulator
towards higher values of a substrate permittivity.}


\begin{document}

\maketitle

\section{INTRODUCTION}

At present there exists a considerable interest
 to the unique electronic properties of graphene \cite {Novoselov:04:1, Novoselov:09:1, Geim:07:1}.
It has been shown that many-body effects play an important role  in the description of physical phenomena in graphene.

 Quasiparticles in graphene interact via Coulomb law with the effective coupling constant
$\alpha_{eff} \sim \alpha_{em}/v_f \sim 300/137 \sim $ 2. Thus the system is strongly coupled and the application of analytical
methods to study the properties of graphene is difficult and numerical simulation is an adequate method for the investigations.

 Below we present the results of the study of the insulator-semimetal phase transition which is rather important for practical applications of graphene.
The influence of the dielectric properties of the substrate,
external fields, structural defects to the conductivity of a
graphene  may shift the position of the phase transition.
 Below we present the study of the conductivity of the monolayer graphene  as a function of
a concentration of defects  in  the effective field model.

Usually graphene is placed on the substrate and the effective coupling
is $\alpha_{eff} = 2 \alpha_{em}/( v_f (\epsilon+1) )$, where $\epsilon$ is the dielectric permittivity of graphene.
The use of substrates with different values of dielectric
permittivity $\epsilon$ alters the interaction between quasiparticles.

It has been shown in \cite{Lahde:09:1,  Hands:08:1,   Buividovich:2012uk} that
a decrease of the substrate  permittivity leads to the shift of the transition of graphene  from  semi-metal phase
to the insulator phase. The phase transition occurs for the values of permittivity $\epsilon_c \sim$ 4.
Recent experimental results show \cite{Mayorov} that graphene is in semi-metal phase even in vacuum ($\epsilon_c =$ 1).  The possible explanation of this discrepancy is a modification (screening) of the Coulomb potential at short distances \cite{Ulybyshev}.  However the phase transition still exists for $\epsilon_c<$ 1  even in the case of screened Coulomb potential. Also qualitative behaviour of chiral condensate, which is the order parameter of the phase transition is quite similar to the effective field theory with unscreened Coulomb potential. So we can hope that influence of defects in effective field theory will describe real physics properly  at qualitative level.

\section{DETAILS OF THE CALCULATIONS}

The Euclidean partition function for the electrons in graphene is calculated using the Monte Carlo method  \cite{Novoselov:04:1, Novoselov:09:1, Geim:07:1, semenoff}
\begin{eqnarray}
\label{partfan}
 \mathcal{Z}   =
 \int \mathcal{D}\bar{\psi} \mathcal{D}\psi \mathcal{D}A_0
 \exp\left( -\frac{1}{2}\int d^4x \lr{\partial_{i} A_{0}}^2
 - \right. \nonumber\\ \left. -
 \int d^3x \, \bar{\psi}_f \, \lr{ \Gamma_0 \, \lr{\partial_0 - i g A_0}
 - \sum\limits_{i=1,2}
    \Gamma_i \partial_i} \psi_f
 \right) ,
\end{eqnarray}
where $A_{0}$ is the zero component of the vector potential of the electromagnetic field, $\Gamma_{\mu}$ are the Euclidean gamma - matrices and $\psi_f$, ($f\hm = 1, 2$) are two flavors of Dirac fermions, corresponding to the two spin components of the electronic excitations in graphene, the effective coupling constant $g^2\hm = 2 \alpha_{em}/( v_F (\epsilon+1))$ ($\hbar\hm =\hm c\hm =1$). The zero component of the vector potential  $A_0$ satisfies the periodic boundary conditions in space and time $A_{0} (t=0)=A(t=1/T)$, where $T$ is the temperature. In the absence of the  magnetic field fermionic fields obey periodic boundary conditions in space and antiperiodic boundary conditions in time $\psi_f(t=0)=-\psi_f(t=1/T)$.
The partition function (\ref{partfan}) does not include the vector part of the potential, $A_i$, since it is suppressed by the Fermi velocity $v_f\sim 1/300$.

For a discretization of the fermionic part of the action in
 (\ref{partfan}) we use the Kogut-Susskind fermions  \cite{MontvayMuenster,
DeGrandDeTarLQCD}. One kind of these fermions in $2\hm + 1$ dimensions corresponds to
the two flavors of ordinary Dirac fermions
\cite{MontvayMuenster, DeGrandDeTarLQCD, Burden:87:1}, making
them particularly suitable for the simulation of the quantum field theory of
graphene. The action for Kogut-Susskind fermions,
interacting with an abelian lattice gauge field has the form
\begin{eqnarray}
\label{lat_act_interact} S_{\Psi}\lrs{\bar{\Psi}_x, \Psi_x, \theta_{x, \, \mu}} =
 \sum\limits_{x, y} \bar{\Psi}_x \, D_{x, y}\lrs{\theta_{x, \, \mu}} \, \Psi_y
= \nonumber \\ =
 \frac{1}{2} \, \sum\limits_{x} \, \delta_{x_3, \, 0} \, \left(  \sum\limits_{\mu=0, 1, 2}
 \bar{\Psi}_x \alpha_{x, \mu} e^{i \theta_{x, \, \mu}} \Psi_{x+\hat{\mu}}
 - \right. \nonumber \\ \left. -
 \sum_{\mu=0, 1, 2}
 \bar{\Psi}_x \alpha_{x, \mu} e^{-i \theta_{x, \, \mu}} \Psi_{x-\hat{\mu}}
 + m {\bar{\Psi}}_x \Psi_x \right) ,
\end{eqnarray}
where lattice coordinates are: $x^{\mu}\hm
= 0\ldots L_{\mu}\hm -1$, and $x^3$ is bounded by the condition $x^3\hm = 0$
(the graphene sheet lie in the plane $x^3=0$),
$\bar{\Psi}_x$ is one-component Grassmann field, $\alpha_{x,
\mu}\hm = (-1)^{x_0 + \ldots + x_{\mu-1}}$, and $\theta_{x, \,
\mu}$ are the variables on the links, lattice analogues of the vector potential $A_{\mu}\lr{x}$ (only $\theta_{x, \,
0}$ is nonzero due to the suppression of magnetic field), $D_{x, y}$
is the Dirac operator for the Kogut-Susskind fermions. $m$ is an artificial mass. It should be introduced in order to make the lattice Dirac operator invertible. We perform calculations for several different values of $m$ and obtain real physical values of all quantities in the limit  $m \rightarrow 0$.

For the discretization of the electromagnetic field in the action (\ref{partfan})
we use the non-compact action for the gauge fields:
\begin{eqnarray}
\label{gauge_lat_act}
 S_g\lrs{\theta_{x, \, \mu}} = \frac{\beta}{2} \, \sum\limits_x \sum\limits^{3}_{i=1}
 \lr{ \theta_{x, \, 0} - \theta_{x + \hat{i}, \, 0} }^2  ,
\end{eqnarray}
where the sum over $x$ is performed over the entire 4D lattice.
The constant  $\beta$ is defined as follows
\begin{eqnarray}
\label{lattice_coupling_constant}
 \beta \equiv \frac{1}{g^2} = \frac{v_F}{4 \pi e^2} \, \frac{\epsilon + 1}{2} ,
\end{eqnarray}
where the factor  $\frac{\epsilon + 1}{2}$ is due to the screening of the electrostatic interactions.

Since the fermion action
(\ref{lat_act_interact}) is bilinear in the fermion fields, it can be integrated out, which gives
\begin{eqnarray}
\label{ferm_integrated}
 \mathcal{Z} =
 \int \mathcal{D}\bar{\Psi}_x \, \mathcal{D}\Psi_x \, \mathcal{D}\theta_{x, \, 0}\,
 \nonumber \\
 \exp\lr{ - S_g\lrs{\theta_{x, \, 0}} - S_{\Psi}\lrs{\bar{\Psi}_x, \Psi_x, \theta_{x, \, 0}}}
 = \nonumber \\ =
 \int \mathcal{D}\theta_{x, \, 0}\,
 \det\lr{ D\lrs{ \theta_{x, \, 0} } }
 \expa{ -S_g\lrs{\theta_{x, \, 0}} }  .
\end{eqnarray}
Thus, we get the effective action
\begin{eqnarray}
\label{eff_action}
 S_{eff}\lrs{ \theta_{x, \, 0} } = S_g\lrs{ \theta_{x, \, 0} } - \ln \det\lr{ D\lrs{ \theta_{x, \, 0} } } ,
\end{eqnarray}
which includes the determinant $\det\lr{ D\lrs{ \theta_{x, \, \mu} }}$ of the Dirac operator $D_{x, y}\lrs{\theta_{x, \,
\mu}}$ (\ref{lat_act_interact}).

For the numerical simulations the standard  hybrid Monte-Carlo method was used, allowing to generate gauge field configurations  $\theta_{x, \, 0}$ with statistical weight $\expa{-S_{eff}\lrs{ \theta_{x, \, 0} }}$ \cite{ MontvayMuenster, DeGrandDeTarLQCD}.

To find the semimetal-insulator phase transition it is convenient to use the order parameter the chiral
fermion condensate $ \vev{ \bar{\psi} \, \psi }$.
In the semi-metal phase the chiral condensate $ \vev{ \bar{\psi} \, \psi }=0$, and in the insulator phase  $ \vev{ \bar{\psi} \, \psi } \neq 0$.

In terms of the Kogut-Susskind fermions the condensate $ \vev{ \bar{\psi} \, \psi }$ has the form:
\begin{eqnarray}
\label{condensate_def}
 \vev{ \bar{\psi} \, \psi }
 =
 \frac{1}{8 \, L_0 \, L_1 \, L_2} \, \sum\limits_{x, t} \, \vev{\bar{\Psi}_x \Psi_x }  .
\end{eqnarray}
After integrating over fermions, the chiral condensate (\ref {condensate_def}) can
be expressed as the average of certain combinations of the fermion propagator $D^{-1}_{x, y}\lrs{\theta_{x, \mu}}$, calculated with the weight (\ref{ferm_integrated}). Also we measure conductivity of graphene sheet. It can be obtained from the value of the
correlator of electromagnetic currents at the central point.

\section{RESULTS}
Below we consider the dependence of the conductivity of monolayer graphene   on the
defect concentration.
 In the effective field-theoretic model defects of
graphene crystal  can be described through the
effective change of the probability of the fermion excitation  hopping
 from site to site.
The absence of the node in a crystalline
lattice means a zero probability of hopping of fermionic excitation to this site.
In the effective graphene theory we simulate the zero probability
of fermion hopping to this node using the vanishing of
the corresponding non-diagonal elements of the matrix of the fermion
operator $D_{x,y}$. The  defects are located
randomly on graphene plane.

\begin{figure}[htb]
  \includegraphics[width=9cm, angle=0]{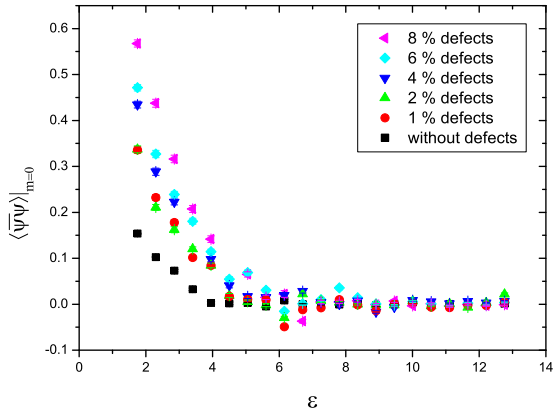}
  \caption{Fig.1. The value of the chiral condensate versus the dielectric
  permittivity of substrate for the different  concentratins of defects.}
  \label{fig1}
\end{figure}

We generate statistically independent field configuration on $20^4$ lattice for
different values  of dielectric permitttivity $\epsilon $  and  percentage of the defects.
For each set of lattice parameters we  calculate the chiral condensate and conductivity.
 The dependence of the fermion condensate on the
dielectric permittivity substrate at various concentrations
defects is shown in Fig.\ref{fig1}.

\begin{figure}[htb]
  \includegraphics[width=9cm, angle=0]{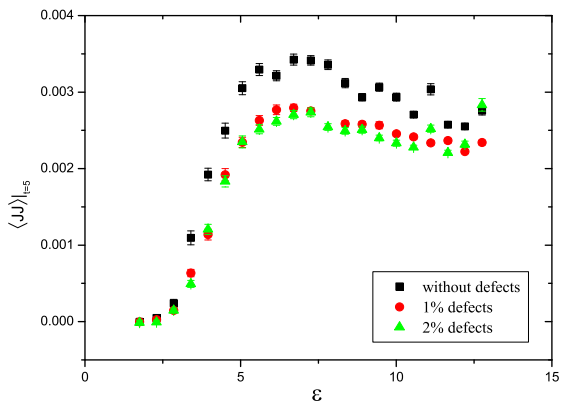}
  \caption{Fig.2. The conductivity of graphene versus the permittivity for different concentrations of defects in
  the substrate.}
  \label{fig2}
\end{figure}
Since the ''smeared'' conductivity \cite{Buividovich:2012uk} is in direct proportion to the correlator of electromagnetic currents at central point and we are interested in the qualitative dependence of conductivity on the interaction strength and defects concentration, we can study the electromagnetic currents correlator.
This dependence is shown in Fig.\ref{fig2}.

\begin{figure}[htb]
  \includegraphics[width=9cm, angle=0]{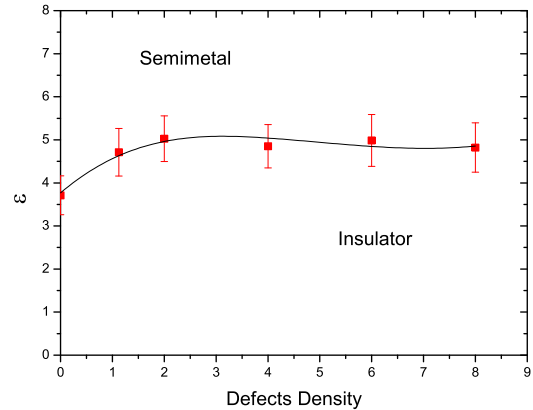}
  \caption{Fig.3. The phase diagram for the semimetal-insulator phase transition
  in the plane defects density - dielectric permittivity of substrate.}
  \label{fig3}
\end{figure}

Fig.\ref{fig1} and \ref{fig2} shows that for certain values of the
substrate  permittivity the strong change of the
conductivity of graphene and the magnitude of the fermion condensate takes place.
 This change indicates the phase transition,
graphene goes from the  state of the conductor to the state of the dielectric.
Using these data we can plot a phase diagramme in a plane the
substrate permittivity - the concentration of defects.

Fig.\ref{fig3} shows a weak dependence of the points of the
phase transition on the concentration of defects in the effective field theory.

\section{Conclusions}

An important conclusion of the analysis is that the presence of defects in the substrate
leads to a shift of the phase transition to higher values of the dielectric permittivity.
It  means that the presence of  defects in graphene may facilitate the transition from state of the conductor
to the  dielectric state.

The authors are  grateful to Mikhail Zubkov for interesting and useful
discussions. The calculations were performed on a cluster of ITEP ``Stakan'',
MV~100K at the Moscow Joint Supercomputer Center and at the
Supercomputing Center of the Moscow State
University.


\end{document}